\documentclass[twoside]{article}
\usepackage{fleqn,espcrc2,epsf}

\usepackage{graphicx}
\usepackage{graphics}
\usepackage{epsfig}
\usepackage[figuresright]{rotating}

\newcommand{\AmS}{{\protect\the\textfont2
  A\kern-.1667em\lower.5ex\hbox{M}\kern-.125emS}}

\title{An Investigation of the Soft Pion Relation in Quenched Lattice QCD}
 \author{UKQCD Collaboration\\
         K.C.~Bowler\address{Department of Physics and Astronomy, 
 University of
          Edinburgh EH9 3JZ, UK}, 
         L.~Del Debbio\address{Department of Physics and Astronomy, 
 University 
         of Southampton, SO17 1BJ, UK},
         J.M.~Flynn$^{\rm b}$,
         L.~Lellouch\address{Theory Division, CERN, CH 1211, Geneva 23, 
         Switzerland},
         V.I.~Lesk$^{\rm b}$\thanks{Presenter},
         J.~Nieves\address{Departamento de Fisica Moderna, Universidad de 
         Granada, Granada 18071, Spain}, 
         C.M.~Maynard$^{\rm a}$ and
         D.G.~Richards$^{\rm a}$\thanks{Jefferson Laboratory, Newport 
 News, 
         VA 23606, USA}.}

\begin{document}

\begin{abstract}
A lattice determination of the form factor and decay constants for
the semileptonic decay of heavy pseudoscalar (PS) mesons at zero
recoil is presented from which the soft 
pion relation is satisfied. Chiral extrapolation of the form factor is
performed at constant $q^2$. Pole dominance is used to extrapolate the
form factor in heavy quark mass. At the B mass, the form factor at
zero recoil lies somewhat below the ratio of decay constants; the
relation remains satisfied within error.
\vspace{-6pt}
\end{abstract}

\maketitle

\section{Introduction}
Heavy-quark and chiral symmetries combine to predict the following
relation for 
a heavy-light meson H the mass of whose heavy quark is larger than some
arbitrary hadronic scale, and for a soft pion:
\begin{equation}
f_0(q^2_{\mathrm{max}})_{H\rightarrow\pi}=f_H/f_{\pi}
\label{equation:spr}
\end{equation}
where $f_0(q^2)$ is the form factor in the $0^+$ channel
for the vector current hadronic matrix element $\langle
H(p)|V^{\mu}|\pi(p-q)\rangle$. The two quantities on the right-hand 
side are the corresponding PS meson decay
constants. Equation~\ref{equation:spr} is the soft pion relation for
heavy-light PS mesons~\cite{neunirligbur}.

Results are given for $f_0(q^2_{\mathrm{max}})_{H\rightarrow\pi}$ and
$f_H/f_\pi$ for heavy-light PS mesons and are extrapolated to the B
mass. Errors are quoted as 
\begin{equation}
f_B=190(5)_{\mathrm{stat.}}(10)_{\mathrm{sys.}} \mathrm{MeV}.
\end{equation}
\section{Extracting the form factors}
The form factors for the hadronic matrix element of semileptonic
heavy pseudoscalar decay to a pion are extracted from mass
extrapolations of the appropriate 
matrix elements of quenched lattice QCD with SW action
~\cite{sw} with $a^{-1}$=2.6(1) GeV, $\beta$=6.2 on a $24^3\times 48$ lattice
using 216 gauge configurations generated with a
combination of the Cabbibo-Marinari algorithm~\cite{cma} and an
over-relaxed algorithm~\cite{oa}. The
matrix 
elements are calculated using the method of extended quark
propagators with all combinations of the following hopping parameters:
\begin{eqnarray*}
\kappa_{\mathrm{heavy}}&\in&\{0.1200, 0.1233, 0.1266, 0.1299\} \\
\kappa_{\mathrm{light}}&\in&\{0.1346, 0.1351, 0.1353\} \\
\kappa_{\mathrm{spectator}}&\in&\{0.1346, 0.1351\}. 
\end{eqnarray*}
From the light PS meson masses $\kappa_{\mathrm{crit}}$=0.13582(1)(2) is
found. 

The action and operators are improved to O($a$) and
renormalized to a continuum scheme, using 
coefficients determined non-perturbatively where
possible~\cite{cswca,zvzabv}. The following values are
\vspace{12pt} used:  
\begin{tabular}{cccccc}	
$Z_V$&=&0.792&$Z_A$&=&0.807\\ 
$b_V$&=&1.41&$b_A$&=&1.11\\
$c_V$&=&$-1.58\times 10^{-2}$&$c_A$&=&$-3.71\times 10^{-2}$\\
$c_{SW}$&=&1.61&$b_M$&=&\vspace{12pt}0.583
\end{tabular}	
Of these, only $c_V$, $b_A$ and $b_M$
are determined from a perturbative scheme~\cite{bacvbm}.

In the continuum, the form factors $f_+$ and $f_0$ are defined from
hadronic matrix elements as follows: \\
$\langle H(p)|V^\mu|\pi(p-q)\rangle= $\vspace{5pt}
\\
$f_+(q^2)\left(2~p^\mu-\left[\left(M_H^2-M_\pi^2\right)/q^2+1\right]q^\mu\right)$+\vspace{5pt}
\\
$f_0(q^2)\left(\left[M_H^2-M_\pi^2\right]/q^2\right)~q^\mu$

\section{Pion physics on the lattice}
Lattice correlators in this study are directly relevant to the physics
of heavy-light mesons with mass $1300-2200$ MeV, and of light-light mesons
with mass $350-850$ MeV. Form factors at the same value of $q^2$ for
decays to a pion can be estimated by assuming the following
dependence of a form factor $f \in \left\{f_+,f_0\right\}$ on the
meson masses: 
\begin{equation}
f(q^2,M,m)=f(q^2,M_{\mathrm{chiral}},0)+a\times\Delta M+b\times m^2
\end{equation}
and fitting the data to determine $f(0)$ and $a$.
Similarly light-light meson and heavy-light meson decay contants 
$f_L$ and $f_H$
are extrapolated in the meson mass according to the following:

\begin{equation}
f_L(m)=f_L(0)+a^{\prime}\times m^2
\end{equation}
\begin{equation}
f_H(M)=f_H(M_{\mathrm{chiral}})+a^{\prime\prime}\times\Delta M
\end{equation}
where $\Delta M\equiv M-M_{\mathrm{chiral}}$.

\begin{figure}[htb]
\centerline{\setlength\epsfxsize{0.9\hsize} \setlength\epsfysize{0.65\hsize}
\vspace{-22pt}
\epsfbox{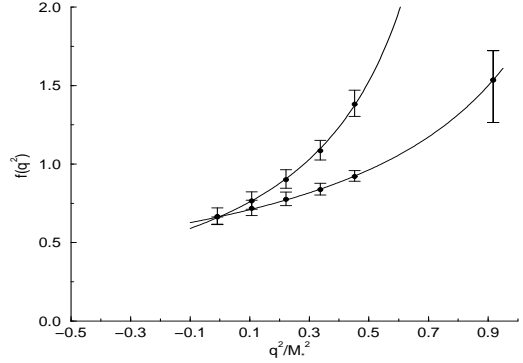}}
\caption[]{Example fit to $f_+,f_0(q^2)$ for decay to a pion,
$\kappa_{\mathrm{heavy}}=0.1200$.}
\label{fig:BKsimh}
\end{figure}

The $q^2$ dependence of the form factors can be modelled using the
following parameterization~\cite{bk}.

\begin{eqnarray}
f_+(q^2)&=&\frac{c_B(1-\alpha)}{(1-q^2/M_{\ast}^2)(1-\alpha q^2/M_{\ast}^2)}\\
f_0(q^2)&=&\frac{c_B(1-\alpha)}{(1-q^2/\beta M_{\ast}^2)}
\label{equation:BK}
\end{eqnarray}
where $M_{\ast}$ is the heavy-light vector meson mass.
$f_0(q^2_{\mathrm{max}})$ is extracted by extrapolating $f_0$ upwards in $q^2$ using
the best fit curve from a fit of ~\ref{equation:BK} ~(fig.~\ref{fig:BKsimh}).
One finds the soft pion relation to be well satisfied at simulated
values of heavy quark mass.
\begin{table}[htb]
\caption{The soft pion relation for a massless pion.}
\label{tab:kh1}	
\begin{tabular}{ccc}	
\hline
$\kappa_{\mathrm{heavy}}$&$f_0(q^2_{\mathrm{max}})$&${f_{PS}}/f_\pi$\\
\hline
0.1200&1.5(2)(3)&1.57(6)(9)\\
0.1233&1.5(2)(2)&1.53(6)(9)\\
0.1266&1.5(2)(2)&1.48(6)(9)\\
0.1299&1.4(1)(1)&1.41(5)(9)\\
\hline
\end{tabular}	
\vspace{-14pt}	
\end{table}
\section{The soft pion relation for $B\rightarrow\pi$}
\begin{figure}[htb]
\centerline{\setlength\epsfxsize{0.9\hsize} \setlength\epsfysize{0.65\hsize}
\vspace{-22pt}
\epsfbox{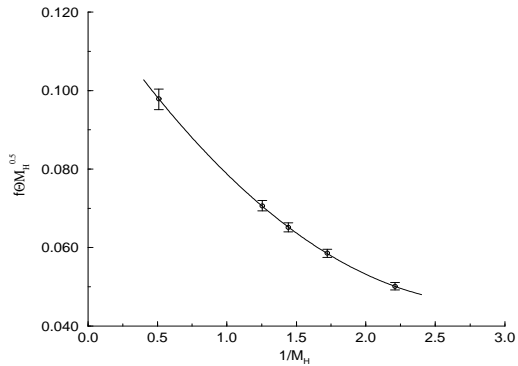}}
\caption[]{Best fit $f_B$ vs. heavy mass
according to eqn.~\ref{equation:eq2}.}
\label{fig:spthe}
\end{figure}

Heavy quark effective theory predicts the following form for the
dependence of $f_0$ on the heavy quark mass, at constant recoil
variable
$v\cdot (p-q)$:
\begin{equation}
f_0(M,\omega)\Theta (M)\sqrt{M}=a+b/M+c/M^2+O({\frac{1}{M^3}})
\label{equation:eq1}
\end{equation}
and for a heavy-light pseudoscalar decay constant:
\begin{equation}
f_H(M)\Theta
(M)\sqrt{M}=a^{\prime}+b^{\prime}/M+c^{\prime}/M^2+O({\frac{1}{M^3}}) 
\label{equation:eq2}
\end{equation}
where $\Theta (M)$ is a perturbative matching coefficient function whose
value is very close to 1.
These prescriptions are used as fit ansatz to determine a best fit
quadratic in 1/M to the data in table~\ref{tab:kh1}. The curve is
shown in (fig.~\ref{fig:spthe}). 
Values for $f_0(q^2_{\mathrm{max}})$ and $f_H$ extrapolated to the b
mass are presented in table~\ref{tab:kh2}.
\begin{table}[htb]
\caption{The soft pion relation for B decay to a massless pion.}
\label{tab:kh2}	
\begin{tabular}{ccc}	
\hline
&$f_0(q^2_{\mathrm{max}})$&${f_B}/f_\pi$\\
\hline
$M_B$&1.1(2)(2)&1.45(6)(9)\\
\hline
\end{tabular}	
\vspace{-14pt}	
\end{table}
It is instructive to repeat the whole procedure, extrapolating form
factors for a pion which is not massless but has its physical
mass. Results are presented in table~\ref{tab:kh3} and are practically
unchanged from the massless case.
\begin{table}[htb]
\caption{The soft pion relation for B decay to a $140$ MeV pion.}
\label{tab:kh3}	
\begin{tabular}{ccc}	
\hline
&$f_0(q^2_{\mathrm{max}})$&${f_B}/f_\pi$\\
\hline
$M_B$&1.1(2)(2)&1.44(6)(9)\\
\hline
\end{tabular}	
\vspace{-14pt}	
\end{table}
\section{Systematic error}
Systematic error in this work arises from:
\begin{enumerate}\itemsep0pt\parsep0pt
\item discretization errors of O($a^2$)
\item quenched approximation
\item estimation of correlation matrix for fits
\item interpolation of lattice form factors in $q^2$
\item corrections to heavy quark effective theory
\item choice in setting the scale.
\end{enumerate}

The quoted systematic error is an estimate of the variation arising
from the last sources $3-6$, achieved by repeating the analysis with
different numbers of bootstrap sets, varying the model function used
to interpolate the form factor in $q^2$, changing the degree of the
polynomial in 1/M used as an ansatz for the extrapolation in heavy
mass, and setting the scale alternatively against $M_\rho$ and the
gluonic scale $r_0$. Quenching error is not quantified here.
Residual discretization errors also may be significant, particularly in
the heavy extrapolation where the fitted curve used
may be diverted substantially by a correction to the
lattice form factor at the largest simulated heavy quark mass,
for which $(Ma)^2\simeq 0.65$. 
\section{Conclusions}
For a simulated heavy-light decay whose mass is within 20\%
of $M_D$, the soft pion relation is reproduced in this study. On
extrapolating the form factor and the heavy-light decay constant to
the B mass, systematic errors become O(30\%), to within which precision
the soft pion relation holds for B meson decays. Other approaches to
date~\cite{JLNRQCD,JLFNALQCD} have generally violated this relation, with
the form factor lying significantly below the ratio of decay constants.
Work is ongoing with the prospect of controlling further the
systematic error, and a simulation at $\beta=6.0$ including a static
heavy quark will address a large source of systematic
uncertainty as well as providing scaling information~\cite{inprep}. 
\bibliography{proc}
\bibliographystyle{unsrt}

\end{document}